\documentclass[a4paper,12pt]{article}
\usepackage{amsmath}
\usepackage{amsfonts}
\usepackage[bookmarksnumbered=true,breaklinks=true]{hyperref}
\usepackage[dvips]{graphicx}
\usepackage{a4}
\usepackage{parskip}
\newcommand{\inv}[1]{\frac{1}{#1}}
\newcommand{\starco}[2]{\left[#1\stackrel{\star}{,}#2\right]}

\newcommand{\co}[2]{\left[#1,#2\right]}
\newcommand{\var}[2]{\frac{\d #1}{\d #2}}

\newcommand{\intx}{\int d^4x}
\newcommand{\intk}{\int d^4k}

\newcommand{\Act}{S}
\newcommand{\ri}{{\rm i}}
\newcommand{\rig}{{\rm i}g}
\newcommand{\e}{\epsilon}

\renewcommand{\d}{\delta}
\renewcommand{\l}{\lambda}
\renewcommand{\th}{\theta}
\renewcommand{\k}{\tilde{k}}
\newcommand{\p}{\tilde{p}}

\newcommand{\bc}{\bar{c}}
\newcommand{\bphi}{\bar{\phi}}

\newcommand{\hd}{\hat{\delta}}

\newcommand{\eth}[1]{\frac{\e_{#1}}{\th}}
\newcommand{\uim}{UV/IR mixing }
\newcommand{\nc}{non-commutative }
\title{\bf A Vector Supersymmetry Killing IR Divergences in Non-Commutative Gauge Theories}
\author{Daniel N.~Blaschke}
\date{October 15, 2007}
\begin{document}
\maketitle
\begin{center}
\renewcommand{\thefootnote}{\fnsymbol{footnote}}
\vspace{-0.3cm}Institute for Theoretical Physics, Vienna University of Technology\\Wiedner Hauptstrasse 8-10, A-1040 Vienna (Austria)\\[0.5cm]
\ttfamily{E-mail: blaschke@hep.itp.tuwien.ac.at}
\vspace{0.5cm}
\end{center}
\begin{abstract}
This is a report on the joint work with Fran\c cois Gieres, Stefan Hohenegger, Olivier Piguet and Manfred Schweda. We consider a \nc U(1) gauge theory with an extension which was originally proposed by A.~A. Slavnov~\cite{Slavnov:2003,Slavnov:2004} in order to get rid of \uim problems. Here we show, that the improved IR behaviour of this model is mainly due to the appearence of a linear vector supersymmetry.
\end{abstract}
\newpage
\tableofcontents

\section{Introduction}
We consider 3+1 dimensional $\th$-deformed Minkowski space-time $\mathbb{M}_{\th}^4$, with the commutation relation
\begin{align}
\co{\hat{x}^\mu}{\hat{x}^\nu}=\ri\th^{\mu\nu}
\end{align}
for the space-time coordinates (cf.~\cite{Snyder:1946, Filk:1996}, see also~\cite{Douglas:2001} for a review). The non-commutativity parameter $\th^{\mu\nu}$ is assumed to be constant and in order to avoid difficulties with time-ordering in the field theory, we choose the special case where $\th^{0\mu}=0$. In order to construct the perturbative field theory formulation, it is more convenient to
use fields $A(x)$ (which are functions of ordinary commuting coordinates) instead of operator valued objects like $\hat{A}(\hat{x})$. One therefore defines the linear map $\hat{f}(\hat{x})\mapsto S[\hat{f}](x)$, called the ``symbol'' of the operator $\hat{f}$. One can then represent the original operator multiplication in terms of star products of symbols as
\begin{align}
\hat{f}\hat{g}=S^{-1}\left[S[\hat{f}]\star S\left[\hat{g}\right]\right].
\end{align}
In using the Weyl-ordered symbol (which corresponds to the Weyl-ordering prescription of the operators) one arrives at the following definition of the Weyl-Moyal $\star$-product ($S[\hat{f}](x)\to A(x)$):
\begin{align}
A_1(x) \star A_2( x ) &= e^{\frac{\ri}{2}\th^{\mu\nu} \partial^x_\mu\partial^y_\nu} A_1(x) A_2(y)\Big|_{x = y}\,.
\end{align}
It has the important property of invariance under cyclic permutations of the integral
\begin{align}
\intx A_1(x) \star A_2(x)\star A_3 ( x )&=\intx A_3 ( x )\star A_1(x) \star A_2(x)\,.
\end{align}
Moreover, bilinear terms are unaffected by the star product:
\begin{align}
\intx A_1(x) \star A_2(x)&=\intx A_1(x) A_2(x)\,.
\end{align}
For a field theory this means that interaction vertices gain phases, whereas propagators remain unchanged. In constructing Feynman graphs one hence has to deal with so-called planar and non-planar diagrams~\cite{Minwalla:1999}. While planar diagrams have the same ultraviolet divergences known from commutative field theory, the non-planar ones are finite due to phase factors. A simple example of an integral appearing in non-planar graphs is
\[
\intk \frac{e^{ik\p}}{k^2+\ri\e}\propto\inv{\p^2}\quad \text{with}\quad\p^\mu=\th^{\mu\nu}p_\nu\,.
\]
It is obvious, that the phases act as UV-regulators, but since this regulating effect can only take place for non-vanishing $\p$, a new infrared divergence appears as $\p\to0$. This is the origin of the \emph{\uim} problem~\cite{Micu:2000,Susskind:2000}.

Here we would like to consider a \nc $U(1)$ gauge theory action
\begin{align}
S&=-\inv{4}\intx F_{\mu\nu}\star F^{\mu\nu}\,,
\end{align}
where the field tensor
\begin{equation}
F_{\mu\nu}=\partial_\mu A_\nu-\partial_\nu A_\mu-\ri g\starco{A_\mu}{A_\nu}
\end{equation}
is endowed with a non-Abelian structure due to the star product. As shown by several authors~\cite{Blaschke:2005b,Hayakawa:1999,Ruiz:2000}, this action leads to an IR singular vacuum polarization, whose quadratic IR divergent term
\begin{align}\label{eq:Pi-IR}
&\Pi^{\mu\nu}_{\text{IR}}(k)
=\frac{2g^2}{\pi^2} \, \frac{\k^\mu\k^\nu}{(\k^2)^2}\quad\text{with}\quad\k^\mu=\th^{\mu\nu}k_\nu
\end{align}
is gauge fixing independent. Obviously, graphs with this insertion are IR divergent.

\section{The Slavnov Term}
\begin{figure}[ht]
\centering\includegraphics[scale=0.75]{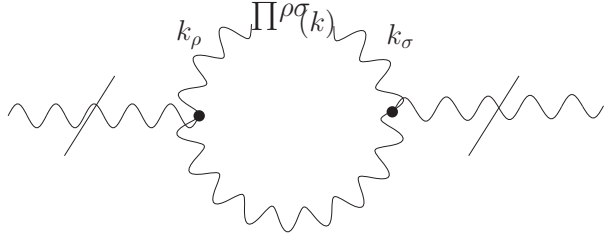}
\caption{this graph has now become \emph{IR finite}}
\label{fig:insertion}
\end{figure}

In order to get rid of IR divergences, Slavnov~\cite{Slavnov:2003,Slavnov:2004} has proposed a modification of Yang-Mills theories, adding to the action a term
\begin{align}\label{sl-term}
\inv{2}\intx \, \l \star\th^{\mu\nu}F_{\mu\nu}.
\end{align}
In doing this, a constraint $\th^{\mu\nu}F_{\mu\nu}=0$ is implemented which has the effect of making the gauge field propagator transversal with respect to $\k^\mu$, i.e. $\k^\mu\Delta^{AA}_{\mu\nu}(k)=0$. Hence, IR divergent Feynman graphs such as the one depicted in Figure~\ref{fig:insertion} become \emph{finite}.

\begin{figure}[ht]
\centering
\includegraphics[scale=0.7]{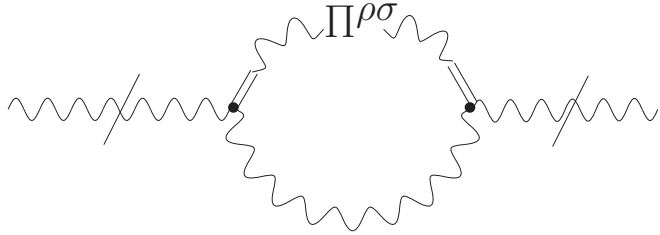}
\caption{example of an IR divergent graph}
\label{fig:insert-2loop}
\end{figure}

There is, however, a catch: Even though $\l$ is introduced as a Lagrange multiplier implementing a constraint, it becomes a dynamical field. Or to be more precise: One has additional Feynman rules, namely a $\l$-propagator, a mixed $\l A$-propagator and a $\l AA$-vertex, and hence numerous additional Feynman graphs. Since the additional propagators are not transversal with respect to $\k^\mu$, Slavnovs trick does not work for certain diagrams, i.e. the one depicted in Figure~\ref{fig:insert-2loop}.

Let us now discuss special features of the gauge fixed action including the Slavnov term: In order to avoid unitarity problems~\cite{Seiberg:2000} we choose the non-commutativity tensor spacelike, i.e.
\[\th^{ij}=\th\e^{ij}\ ,\quad i,j=1,2\]
Furthermore, we choose the gauge fixing to be of an axial type~\cite{Kummer:1961, Schweda-book:1998} with gauge fixing vector in the plane of the non-commutative coordinates:
\[n^I=0\ ,\quad I=0,3\]
With these choices the Slavnov term, together with the gauge fixing terms, have the form of a 2-dimensional topological BF model (cf.~\cite{Blaschke:2006a,Blaschke:2005a} and references therein):
\begin{align}
S_{{\rm inv}} = \intx \ ( -\frac{1}{4}F_{\mu\nu}\star F^{\mu\nu}
+ \frac{\th}{2} \l \star \, \e^{ij}F_{ij} )
\end{align}
and
\begin{align}
S_{{\rm gf}} = \intx \ ( B\star n^iA_i-\bc\star n^iD_ic )
\end{align}
with the covariant derivative
\[D_{\mu} c  = \partial_{\mu} c -\ri g\starco{A_{\mu}}{c}\,.\]

\section{Symmetries \& Consequences}
The action $S=S_{{\rm inv}}+S_{{\rm gf}}$ is invariant under the BRST symmetry
\begin{align}
&sA_\mu = D_\mu c \, ,  &&s \bar c  = B \, , \nonumber\\
 &s\l  = - \ri g \, [\l, c ] \, ,  &&sB = 0 \, , \nonumber\\
&sc  =   \frac{\ri g}{2}  \, [c , c ] \, , &&s^2=0\,,
\end{align}
where we have omitted the stars, and the commutators are considered to be graded by the ghost-number. Additionally, the gauge fixed action is also invariant under a (non-physical) \emph{linear vector supersymmetry (VSUSY)}, whose field transformations are
\begin{align}
&\d_iA_\mu = 0 \, , &&\d_ic=A_i \, ,\nonumber\\
&\d_i\bc=0 \, ,&&\d_iB=\partial_i\bc \, ,\nonumber\\
&\d_i\l  = \frac{ \e_{ij} }{\th} n^j\bc \, , &&\d^2=0\,.
\end{align}
Since the operator $\d_i$ lowers the ghost-number by one unit, it represents  an antiderivation (very much like the BRST operator $s$ which raises the  ghost-number by one unit).

\textbf{Note:} \emph{Only the interplay of appropriate choices for $\th^{\mu\nu}$ and $n^\mu$ lead to the existence of the VSUSY.}

In contrast to the pure topological theories, we have an additional vectorial symmetry:
\begin{align}
&\hd_iA_J = -F_{iJ} \, , \qquad\hd_i\l=-\eth{ij} D_K F^{Kj} \, ,\nonumber\\
&\hd_i\Phi = 0 \qquad \quad \mbox{for all other fields} \,.
\end{align}
This further symmetry is in fact a (non-linear) symmetry of the gauge invariant action. Its existence is due to the presence of the Yang-Mills part of the action which, in contrast to the BF-type part of the action, involves also $A_0$ and $A_3$. Notice that the algebra involving $s$, $\d_i$, $\hd_i$ and the ($x_1,x_2$)-plane translation generator $\partial_i$ closes on-shell (cf.~\cite{Blaschke:2006a}).

\begin{figure}[h]
\centering
\includegraphics[scale=0.8]{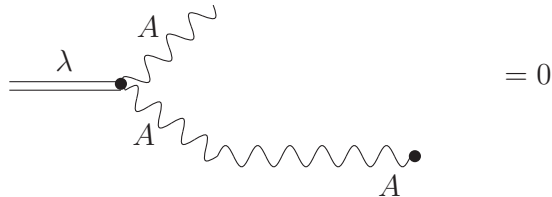}
\caption{The $\l AA$-vertex contracted with a photon propagator vanishes.}
\label{fig:vertex-prop}
\end{figure}

We shall now discuss the consequences of the linear VSUSY: The generating functional $Z^c$ of the connected Green functions is given by the Legendre transform of the generating functional $\Gamma$ of the one-particle irreducible Green functions. At the classical level (tree graph approximation) one has $\Gamma\sim S$, and hence the Ward identity describing the linear vector supersymmetry in terms of $Z^c$ in the tree graph approximation is given by
\begin{align}
\mathcal{W}_i Z^c=\int d^4x \,\Big\{j_B \,\partial_i \var{Z^c}{j_{\bc}} - j_c \,\var{Z^c}{j_A^i} \,
+ \eth{ij} n^j  j_{\l} \, \var{Z^c}{j_{\bc}} \, \Big\}=0\,.
\end{align}
where $\{j_A^\mu,j_\l,j_B,j_c,j_{\bc}\}$ are sources of $\{A_\mu,\l,B,c,\bc\}$, respectively. Varying this expression with respect to $j_c$ and $j^\mu_A$ yields for the gauge field propagator:
\begin{align}\label{prop-relation}
\text{\large\centering\framebox[4cm]{$\Delta_{A_iA_\mu}=0$}}
\end{align}
In other words, as soon as one of its indices is either 1 or 2, the gauge field propagator is zero. As the $\l AA$-vertex is proportional to $\th_{ij}$, which here is non-vanishing only in the ($x_1,x_2$)-plane, relation (\ref{prop-relation}) has the following important consequence for the Feynman graphs: \emph{The combination of gauge boson propagator and $\l AA$ vertex is zero} (see Figure~\ref{fig:vertex-prop}).
Furthermore, it is impossible to construct a closed loop including a $\l AA$-vertex without having such a combination somewhere. \emph{Hence, all loop graphs involving the $\l AA$-vertex vanish!}

In particular, dangerous vacuum polarization insertions as in Figure~\ref{fig:insert-2loop} vanish. This is the reason, why the model is free of the most dangerous, i.e. the quadratic, infrared singularities, as pointed out by Slavnov~\cite{Slavnov:2004} for the special case of $n^\mu=(0,1,0,0)$.

\section{Generalization}
In this section we would like to discuss the question whether we can show cancellation of IR singular Feynman graphs for a more general choice of $\theta^{\mu\nu}$ and $n^\mu$. The answer is yes, but we need to impose stronger Slavnov constraints. The initial Slavnov constraint was $\theta^{12}F_{12}+\theta^{13}F_{13}+\theta^{23}F_{23}=0$. With ``stronger'' we mean that each term in the sum should vanish seperately. Upon imposing these stronger conditions we may write for the action~\cite{Blaschke:2007a}:
\begin{align}\label{3bf-action}
\Act_{\text{inv}}&=\intx\left[-\inv{4}F_{\mu\nu} F^{\mu\nu}+ \inv{2}\e^{ijk}F_{ij}\l_k\right],
\end{align}
with $i,j,k\in\{1,2,3\}$. This action looks like a 3 dimensional BF model coupled to Maxwell theory. As in the pure BF-case, the action has two gauge symmetries
\begin{align}
&\d_{g1}A_\mu=D_\mu\Lambda\,,\qquad&& \d_{g2}A_\mu=0,\nonumber\\
&\d_{g1}\l_k=-\rig\co{\l_k}{\Lambda}\,,\qquad&& \d_{g2}\l_k=D_k\Lambda'\,.
\end{align}
Similar to the previous model, we have an additional bosonic vector symmetry of the gauge invariant action:
\begin{align}
&\hd_iA_0 = -F_{i0} \,, \qquad\hd_i\l_j=\e_{ijk} D_0 F^{0k} \, ,\nonumber\\
&\hd_iA_i = 0 \,.
\end{align}
There is, however, a difference to the previous case: The additional vectorial symmetry is broken when fixing the second gauge symmetry $\d_{g2}$.

If we consider a space-like axial gauge fixing of the form\footnote{$d'=d-\ri g\co{\bphi}{c}$ is the redefined multiplier field fixing the second gauge freedom $\d_{g2}$.}
\begin{align}
S_{{\rm gf}} = \intx \left[ B n^iA_i+d'n^i\l_i-\bc n^iD_ic-\bphi n^iD_i\phi\right],
\end{align}
the gauge fixed action is invariant under the linear VSUSY
\begin{align}
&\d_ic=A_i \, , \qquad\d_i\l_j=-\e_{ijk} n^k \bc\, ,\nonumber\\
&\d_iB=\partial_i\bc \, ,\nonumber\\
&\d_i\Phi=0\,,\qquad\text{for all other fields},
\end{align}
in addition to the usual BRST invariance. The Ward identity describing the linear vector supersymmetry in terms of $Z^c$ at the classical level is given by
\begin{align}
\mathcal{W}_iZ^c=\int d^4x\Big[& j_B\partial_i\var{Z^c}{j_{\bc}}-j_{c} \var{Z^c}{j_A^i} +\e_{ijk}n^j j_\l^k\var{Z^c}{j_{\bc}}\Big]=0.
\end{align}
Hence, the same arguments as before show the absence of IR singular graphs. However, the model exhibits numerous further symmetries which have been discussed in~\cite{Blaschke:2007a}.

One should also note, that a generalization to higher dimensional models is possible. For example if $\l$ had $n$ indices the VSUSY would become
\begin{align}
&\d_ic=A_i\,, &&\d_i\l_{j_1\cdots j_n}=\e_{ikj_1\cdots j_n}n^k\bc\,,\nonumber\\
&\d_iB=\partial_i\bc\,,
\end{align}
after appropriate redefinitions of Lagrange multipliers.

\section{Conclusion and Outlook}
Slavnov-extended Yang Mills theory can be shown to be free of the worst infrared singularities, if the Slavnov term is of BF-type. Furthermore, supersymmetry, in the form of VSUSY, seems to play a decisive role in theories which are not Poincar\'e supersymmetric. Open questions are:
\begin{itemize}
\item What is the role of VSUSY with respect to \uim in topological NCGFT in general?
\item What are the consequences of the additional symmetries appearing in these models?
\end{itemize}

\section*{Acknowledgments}
The author would like to thank S. Hohenegger for helpful comments.\\
D.~N. Blaschke is a recipient of a DOC-fellowship of the Austrian Academy of Sciences at the Institute for Theoretical Physics at Vienna University of Technology.


\end{document}